\documentclass[aps,prl,twocolumn,groupedaddress]{revtex4}

\usepackage{graphicx} 
\usepackage{dcolumn}
\usepackage{bm}

\begin{document}

\title{Magnetic studies of multi-walled carbon nanotube mats: Evidence
for the paramagnetic Meissner effect} 
\author{Guo-meng Zhao$^{*}$ and Pieder Beeli} 
\affiliation{Department of Physics and Astronomy, 
California State University, Los Angeles, CA 90032, USA}
\begin{abstract}
     We report magnetic measurements up to 1200 K on multi-walled
    carbon nanotube mats using Quantum Design vibrating sample
    magnetometer.  Extensive magnetic data consistently show two
    ferrromagnetic-like transitions at about 1000 K and 1275 K,
    respectively. The lower transition at about 1000 K is associated with 
    an Fe impurity phase and its saturation magnetization is in quantitative
    agreement with the Fe concentration measured by an inductively 
    coupled plasma mass spectrometer. On the other hand, the saturation
    magnetization for the higher transition phase ($\geq$1.0 emu/g) is 
    about four 
    orders of magnitude larger than that expected from the measured concentration
    of Co or CoFe, which has a high enough Curie temperature to explain this high transition.  We show that this 
    transition at about 1275 K is not consistent with a magnetic proximity effect
    of Fe-carbon systems
    and ferromagnetism of any carbon-based materials or magnetic impurities but with the paramagnetic
    Meissner effect due to the existence of $\pi$ Josephson junctions in a
    granular
    superconductor.

\end{abstract}
\maketitle 

There are reports of weak ferromagnetism in graphite and carbon-based materials well above room temperature
    \cite{Mend,Kop00,Maple,Mur,Maka}, as well as a theoretical prediction of a ferromagnetic
    instability in graphene sheets \cite{Bas}. It is unclear whether the  high-temperature ferromagnetism is intrinsic or simply 
    caused by contamination of magnetic impurities \cite{Dzw}. There are also
    several reports of high-temperature superconductivity in carbon
    films \cite{Anto,Leb}, carbon nanotubes \cite{Zhao,Tang,Take}, and  graphite
    \cite{Silva,Kop00,Maple}. Gonzalez
    {\em et al.} \cite{Gon} show that both high-temperature ferromagnetic and $p$-wave superconducting instabilities 
    can occur in defective regions of graphite, where topological disorder enhances the density of states.  Schrieffer  
    \cite{Schr} predicts
    ultra-high temperature superconductivity at a quantum critical point
    where ferromagnetic fluctuations are the strongest.
    
    Here we report magnetic measurements up to 1200 K on multi-walled
carbon nanotube mats.  our extensive magnetic data consistently
show two ferrromagnetic-like transitions at about 1000 K
and 1275 K, respectively. We show that the lower transition transition
at about 1000 K is associated with an Fe impurity
phase while the transition at about 1275 K is not consistent
with a magnetic proximity effect in Fe-carbon systems and ferromagnetism of any carbon-based materials
or magnetic impurities but with the paramagnetic Meissner
effect due to the existence of $\pi$ Josephson junctions
in a granular superconductor.

Purified multi-walled
nanotube (MWNT) mat samples are obtained from SES Research of Houston.  
Two different 
samples (Lot $\#$'s RS0656 and RS0657) were prepared by chemical vapor deposition using 
an iron 
catalyst.  By burning off carbon-based materials in air, we find the weights of the residuals to be 2.25$\%$
and 1.725$\%$ for RS0656 and RS0657, respectively.  On the assumption
that the residual contains Fe$_{2}$O$_{3}$, Co$_{m}$O$_{n}$, and
Ni$_{p}$O$_{q}$ (where $m$, $n$, $p$, and $q$ are integers), we determine the relative
metal 
concentrations of the residual using a Perkin-Elmer Elan-DRCe inductively coupled plasma mass spectrometer
(ICP-MS). Since the Co and Ni
concentrations are negligibly small, the relative metal contents are
nearly independent of the valences of Co and Ni we choose for their oxides. From the ICP-MS result and 
the weight of the residual, we obtain the metal-based magnetic impurity 
concentrations (ppm in weight) for RS0656: Fe = 5342.4, Co = 0.5, Ni = 
13.7; and for RS0657: Fe = 6940.8, Co = 36.1, Ni = 20.5.

\begin{figure}[htb]
	 \includegraphics[height=5cm]{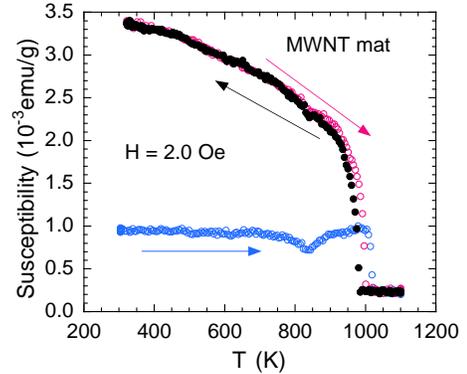}
 \caption[~]{Temperature dependence of the susceptibility in a field of 
2 Oe for a virgin MWNT mat sample (RS0657). }
\end{figure}

Magnetization was measured by a Quantum Design vibrating sample
magnetometer (VSM). The absolute
uncertainty of the temperature is less than 10 K, as checked by the
Curie temperatures of both Ni and Fe. Fig.~1 shows the temperature dependence of the susceptibility in a field of 
2 Oe for a virgin MWNT mat sample (RS0657).   This virgin sample was inserted into the sample chamber
without going through the linear motor used for vibrating 
the sample. A 2 Oe field (using the ultra-low field option) was then applied after the sample was inserted.  From the 
warming data, we clearly see a dip feature at about 833 K, which is 
caused by the competing effect of the ferrimagnetic transition at about
860 K 
for the Fe$_{3}$O$_{4}$ impurity phase and the decomposition of Fe$_{3}$O$_{4}$ into 
the higher Curie temperature 
$\alpha$-Fe phase due to the high vacuum inside the sample chamber (better than
9$\times$10$^{-6}$ torr). From 
the subsequent cooling and warming data, we see that the Curie
temperature ($T_{C}$) of the Fe impurity is about 1000 K, which is
lower than that (1047 K) for the bulk pure $\alpha$-Fe. This is possibly due to 
the doping of carbon into Fe, which can significantly lower 
the $T_{C}$ value \cite{Terry}.  

From Figure 1, it is also apparent that the substantial low-field susceptibility 
(about 2.3$\times$10$^{-4}$ emu/g) persists up to 1100 K, which could arise 
from ferromagnetism of the Co or CoFe impurity phase. It is interesting that
the same magnitude of the paramagnetic susceptibility is also seen in 
sample RS0656 where the Co concentration is only 0.5 ppm. The concentration of Co 
(0.5 ppm) or CoFe (about 1 ppm) is {\em over
three orders of magnitude} too small to explain the measured
paramagnetic susceptibility. Therefore, the observed large paramagnetic susceptibility
well above the $T_{C}$ of the Fe impurity phase should originate from 
ferromagnetism of a carbon-based phase or from the paramagnetic Meissner
effect due to superconductivity.

\begin{figure}[htb]
	 \includegraphics[height=5cm]{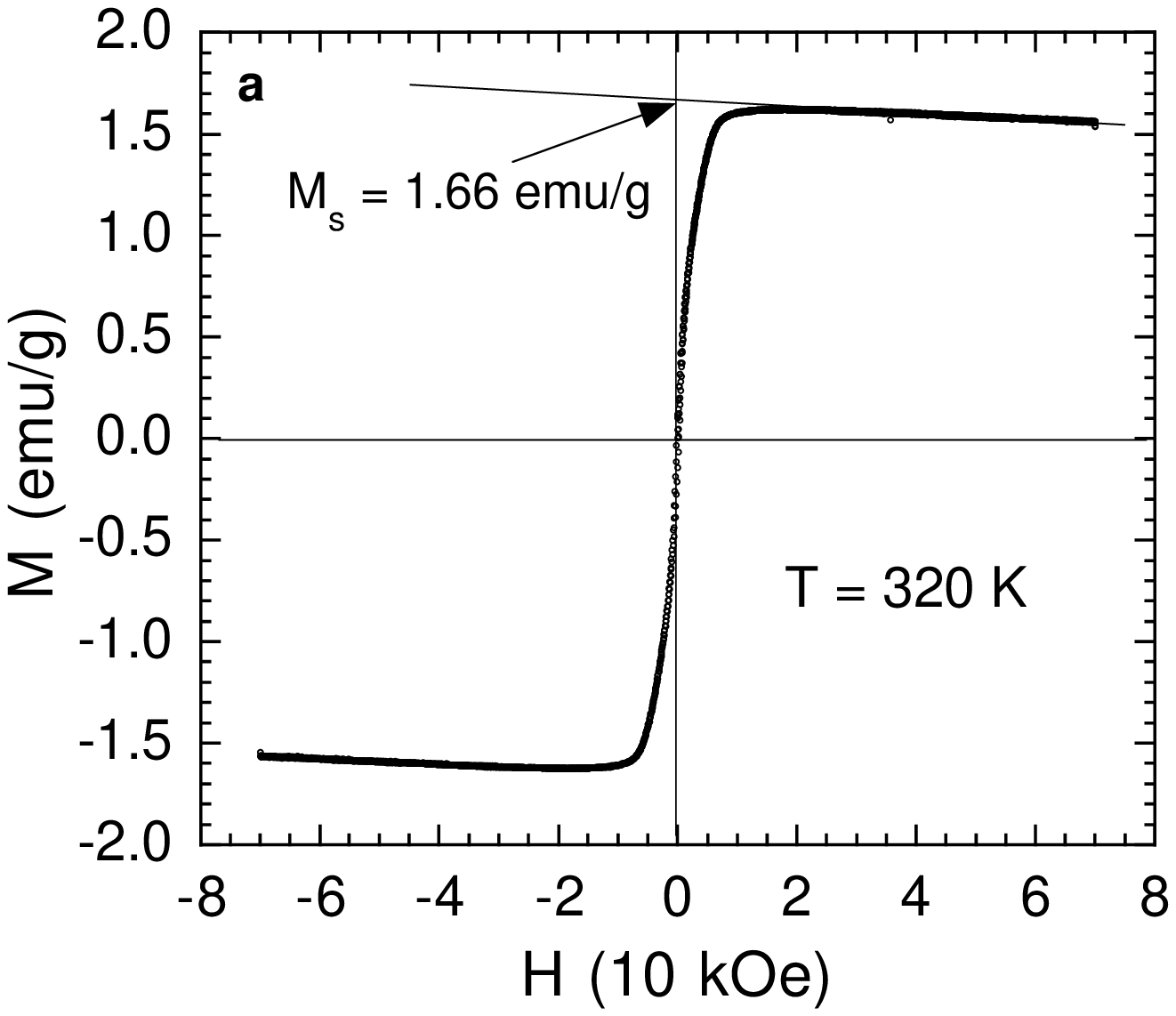}
	 \includegraphics[height=5cm]{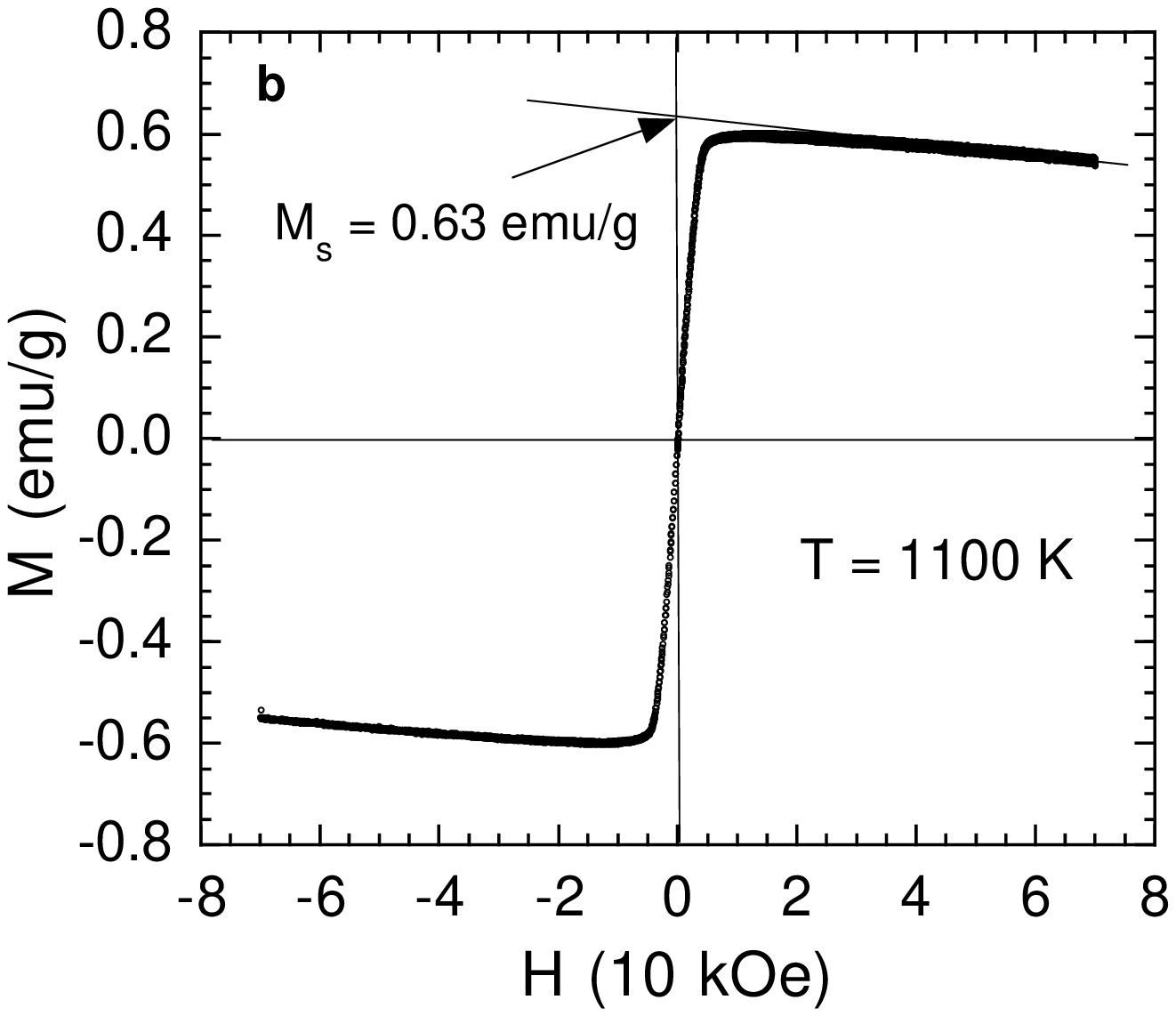}
 \caption[~]{Magnetic hysteresis loops at 320 K (a) and 1100 K
 (b) for sample RS0657.
The coercivity $H_{C}$ is about 140 Oe at 320 K and negligibly small at 1100 K. }
\end{figure}

In Figure 2, we plot the magnetic hysteresis loops at 320 K and 1100
K, which were measured after the above low-field measurements. The coercivity $H_{c}$
is about 140 Oe at 320 K 
while it becomes negligibly small at 1100 K. The saturation
magnetizations $M_{s}$ at 320 K and 1100 K are 1.66 emu/g and 0.63
emu/g, respectively. It is clear that the saturation magnetization at 
1100 K is substantial, indicating a second  magnetic
transition above 1100 K. This second ultra-high temperature ferromagnetic-like
(UHTFL) phase cannot be associated with the
Co (or CoFe) impurity whose concentration is only 36.1 ppm (or 72 ppm) in this sample. Such a small Co 
or CoFe impurity
concentration can contribute a saturation 
magnetization of $<$1.0$\times$10$^{-2}$ emu/g (see~\cite{note}), which is 
about two orders of magnitude
smaller than the measured value (0.63 emu/g). For sample RS0656, which
contains 5342.4
ppm Fe impurity and 0.5
ppm Co impurity, the $M_{s}$ values at 320 K and 1100 K are 1.60 emu/g and 0.59
emu/g, respectively, which are very similar to those for sample RS0657.
The measured $M_{s}$ at 1100 K for sample RS0656 is about {four orders of magnitude}
larger than the value expected from the measured Co or CoFe impurity
concentration. This definitively excludes the Co or CoFe impurity from being the origin of the UHTFL phase.

\begin{figure}[htb]
	 \includegraphics[height=5cm]{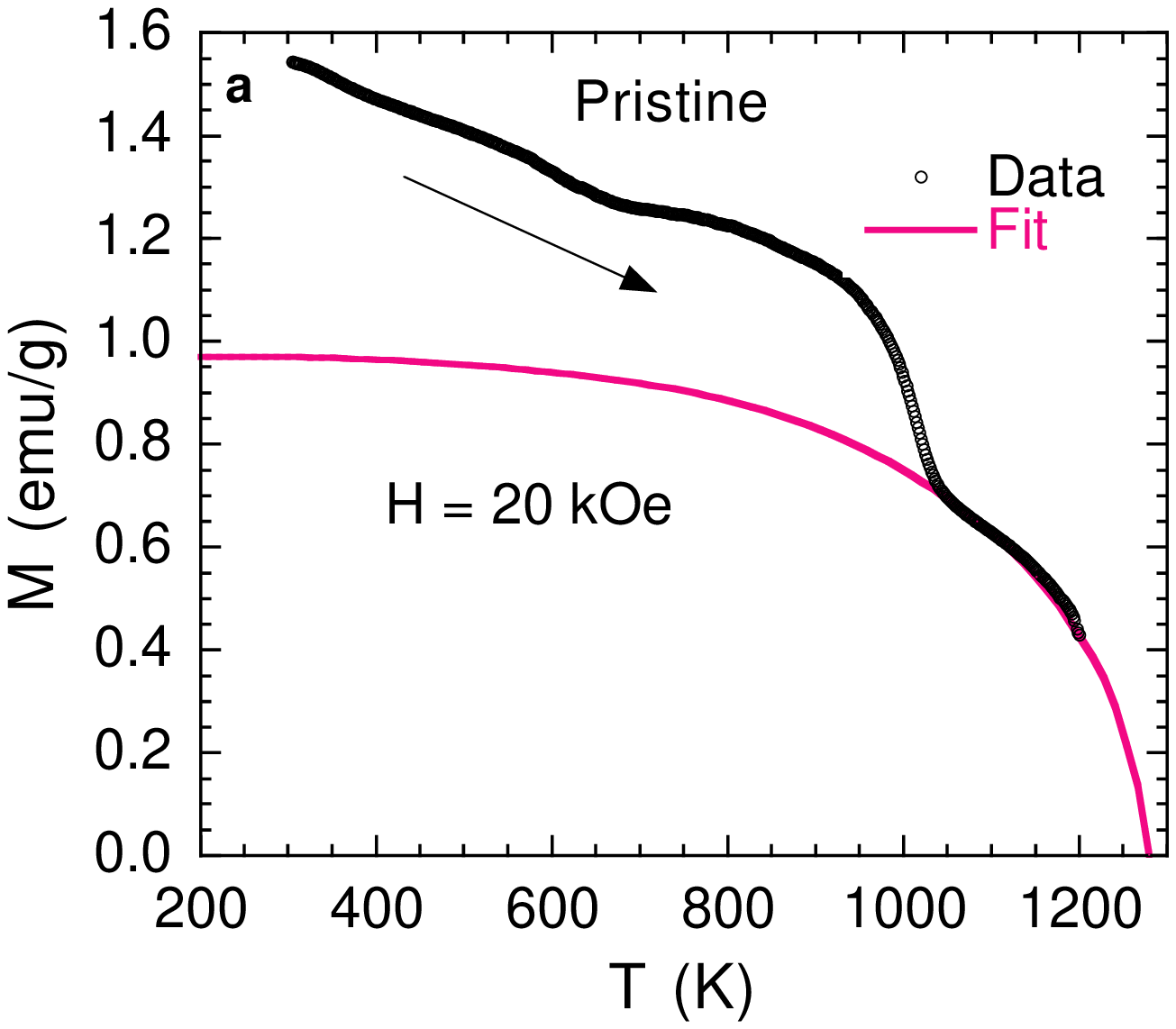}
	 \includegraphics[height=5cm]{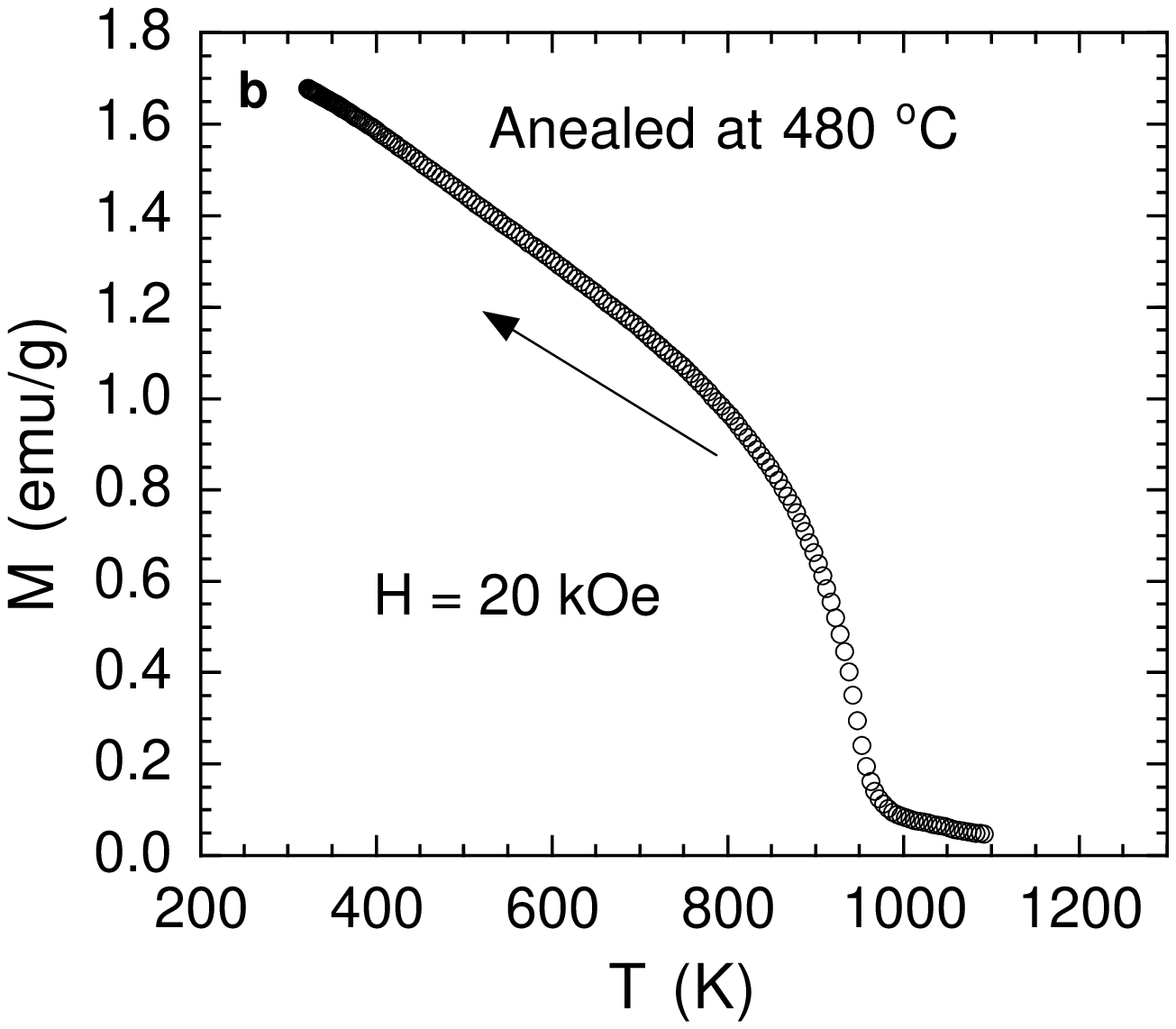}
 \caption[~]{a) Temperature dependence of the magnetization in a field of 
20 kOe for a virgin MWNT mat sample (RS0657). The temperature dependence of the magnetization in 20
kOe should be similar to that of the saturation magnetization
($M_{s}$). The solid line is a fit 
using the curve of $M_{s}(T)/M_{s}(0)$ versus $T/T_{C}$ for Ni,
appropriately scaled to $T_{C}$ = 1275 K. b) Temperature dependence of the magnetization in a field of 
20 kOe for a thermally annealed sample of RS0657. This sample was annealed in air at 480 $^{\circ}$C for 
about 5 minutes.} 
\end{figure}

Since the magnetization in 20 kOe is close to the saturation
magnetization (see Fig.~2), the temperature dependence of the
saturation magnetization can be simulated by the temperature
dependence of the magnetization in 20 kOe. In
Fig.~3a, we plot the temperature dependence of the magnetization in a field of 
20 kOe for another virgin sample of RS0657. One can clearly see that the magnetization above the Curie 
temperature of the Fe impurity phase is large up to 1200 K, implying that the transition 
temperature of the UHTFL phase is higher than 1200 K. If we assume that
the curve of $M_{s}(T)/M_{s}(0)$ versus $T/T_{C}$ for this UHTFL phase is 
the same as that for Ni (see the solid line), we find $T_{C}$ and $M_{s}(0)$ to be 1275 K and 
0.97 emu/g respectively. Then the saturation magnetization at 320 K
contributed from the Fe
impurity phase is 0.57 emu/g.

In Figure 3b, we show the temperature dependence of the magnetization in a field of 
20 kOe for a thermally annealed sample of RS0657. This sample was annealed in air at 480 $^{\circ}$C for 
about 5 minutes and its mass was measured about 1 hour after it was cooled down to room
temperature. The mass of the annealed sample was found to be smaller than that of the pristine
sample by 2$\%$. This mass decrease may be due to the removal of amorphous 
carbon and/or the outershells of MWNTs. The data were taken after an
$M$-$H$ loop was measured at 1100 K
so that all the Fe oxides had been converted to $\alpha$-Fe in such a high
vacuum and temperature.
It is remarkable that the magnetization at 1100 K is reduced by one order of magnitude compared with
that for the pristine sample (see Fig.~3a) while the magnetization at 
320 K increases by 10$\%$.

Figure 4 shows the temperature dependence of the magnetization in a field of 
20 kOe for the residual of sample RS0657, which was obtained by
burning off carbon-based materials in air at 550 $^{\circ}$C for about
10 minutes. The data were similarly taken after an $M$-$H$ loop was measured at 1100 
K.  The specific magnetization was calculated using the mass of the pristine MWNT mat sample.
As indicated by the arrow, the ferromagnetic transition temperature is about 1037 K which is the same as the $T_{C}$ of $\alpha$-Fe if one
considers a thermal lag of about 10 K.  It is striking that the magnetization 
of the residual at 320 K is three times smaller than that for the
annealed sample containing MWNT mats (Fig.~3b). This implies a giant enhancement
of the magnetization of the Fe impurity due to the proximity to MWNTs.
Moreover, we have independently shown that  
the Fe-based impurity phase in virgin samples is Fe$_{3}$O$_{4}$. 
If only this Fe$_{3}$O$_{4}$ impurity phase
is magnetic, the upper limit of the 
room-temperature $M_{s}$ value for sample RS0656 
is calculated to be 0.68 emu/g using the room-temperature $M_{s}$ value (92
emu/g) of the bulk Fe$_{3}$O$_{4}$ and the inferred Fe$_{3}$O$_{4}$ concentration (7372.5 ppm) from 
the metal-based Fe concentration (5342.4 ppm). This upper limit of the
room-temperature $M_{s}$ value due to the Fe$_{3}$O$_{4}$ impurity phase
is about one third of the measured value (1.60 emu/g). It is
interesting that such a giant enhancement of the moment was also observed in one of the Canyon Diablo 
graphite nodule samples (see the results for sample 1.1 of Ref.~\cite{coey}).

\begin{figure}[htb]
	 \includegraphics[height=5cm]{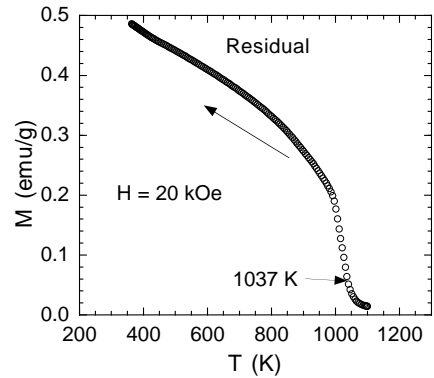}
 \caption[~]{Temperature dependence of the magnetization in a field of 
20 kOe for the residual of sample RS0657, which was obtained by
burning off carbon-based materials in air at 550 $^{\circ}$C for about
10 minutes.}
\end{figure}

From $M$-$H$ loop measurements of the residual, 
we find the saturation magnetizations at 320 K and 1100 K to be 0.52 emu/g and
4.5$\times$10$^{-3}$ emu/g, respectively.  The $M_{s}$ value of the $\alpha$-Fe
impurity at 320 K 
inferred from the data of the residual is in excellent agreement with
that (0.57 emu/g) inferred from the data of the pristine sample
(Fig.~3a). If we calculate $M_{s}$ using the mass of the Fe impurity in
the residual, we find $M_{s}$ to be 74.9 emu per gram of Fe in
agreement with the value used in Ref.~\cite{coey}. Our magnetic measurements on 
Fe$_{3}$O$_{4}$ nanoparticles 
with an average diameter of 40-60 nm also show that $M_{s}$ = 77.0 emu per
gram of Fe at 320 K after the decomposition of Fe$_{3}$O$_{4}$ into Fe
through thermal cycling up to 1100 K in high vacuum. This implies that the $M_{s}$ value of Fe
nanoparticles is significantly smaller than the bulk value ($\sim$200 
emu/g), which could be caused by spin disorder 
on the surface of nanoparticles \cite{Res}. Moreover, the measured $M_{s}$ value at 1100 K 
(4.5$\times$10$^{-3}$ emu/g) is nearly the same as
the $M_{s}$ value at 1100 K (4.5$\times$10$^{-3}$ emu/g) expected from the measured Co
impurity concentration (36.1 ppm) \cite{note}. Such excellent consistencies
between the magnetic data of the residual and the ICP-MS results
further demonstrate that the UHTFL phase seen in Fig.~1, Fig.~2b, and Fig.~3a is not associated with any
magnetic impurity phase.

Now we discuss the origin of the UHTFL phase and the giant enhancement of the
magnetization of the Fe impurity. The UHTFL phase and related giant
moment enhancement of the Fe impurity could arise from magnetic proximity 
between Fe nanoparticles and MWNTs \cite{coey}. However, this picture 
cannot consistently explain the results shown in Fig.~3 and Fig.~4.
Annealing the sample in 480 $^{\circ}$C may destroy the proximity
effect so that the UHTFL phase is converted to the isolated Fe
impurity phase. The converted Fe phase should have a much smaller
saturation magnetization 
than the coupled Fe-MWNT phase which contains extra proximity-induced moments in MWNTs.
The fact that no reduction of the $M_{s}$ value at 320 K is found in
the annealed sample argues against this interpretation. Further, the observed 90$\%$ decrease in the saturation magnetization at
1100 K implies that the coupled Fe-MWNT phase in the annealed sample 
would have been destroyed
by about 90$\%$ so that the $M_{s}$ value at 320 K for the annealed
sample would
be slightly larger than that for the residual sample with no proximity
effect.
However, the $M_{s}$ value at 320 K for the annealed sample (Fig.~3b) is over three
times larger than that for the residual sample (Fig.~4), which makes the above
interpretation very unlikely. 
Another possibility is that the UHTFL phase is related to ferromagnetism of an unknown 
carbon-based material (e.g., amorphous carbon). Such a high Curie temperature
(1275 K) for a carbon-based material would be very remarkable. Annealing the sample in 
air up to 480 $^{\circ}$C removes amorphous carbon,
and the observed 90$\%$ decrease in the saturation magnetization at
1100 K appears to indicate that
the UHTFL phase could be related to amorphous carbon. However, if this interpretation 
were correct, the $M_{s}$ value at 320 K would also be
reduced to about 0.6 emu/g, in sharp contrast to the
measured value (1.7 emu/g). A third possibility is that MWNTs are ferromagnetic with
an ultra-high Curie temperature (1275 K). This is also very unlikely
unless, per Fig.~3b, the annealing procedure causes the Curie temperature of
the MWNTs to be exactly the same as that of the Fe impurity phase.

It is clear that one cannot consistently explain our present data based on 
the magnetic proximity effect and the ultra-high temperature ferromagnetism of unknown carbon-based
phases, MWNTs, or magnetic impurities. Alternatively, if MWNTs are superconducting, 
a MWNT mat 
should be a granular
superconductor. The existence of magnetic impurities in the Josephson
network should lead to the formation of $\pi$ junctions with a negative 
Josephson coupling energy $J$~\cite{Spi}. If the critical current is large
enough, an odd number of $\pi$ junctions within a 
loop generates a spontaneous orbital moment associated with the circulation 
current around the loop \cite{Kus,Dom}. The interaction of these
orbital moments can lead to ferromagnetic-like ordering (orbital
ferromagnetism)\cite{Kus}. Since the
diameter of nanotubes is comparable to the magnetic penetration depth 
\cite{Zhao}, the
diamagnetic Meissner effect is negligibly small so that magnetic field can
enter into the Josephson-coupled network even in a zero-field-cooled
condition. Because the Josephson-coupled loops in MWNT mats should be
much smaller than those in the network made of
conventional superconductors or cuprate superconductors and because
orbital moments are inversely proportional to loop areas \cite{Kus}, the orbital
moments in MWNT mats should be significantly larger. Since $\pi$ junctions can
be formed even if the impurities are not in the ferromagnetic state
\cite{Spi}, the
ferromagnetic-like ordering of the orbital moments can occur above the
Curie temperature of the magnetic impurities. This can explain the
data shown in Fig.~3a. On the other hand, if the Josephson coupling and
critical current are substantially reduced by the removal of the outershells of MWNTs 
after annealing in air at 480 $^{\circ}$C, the density of the $\pi$ junctions 
and thus the orbital moments
are significantly reduced above $T_{C}$ of the magnetic impurities. However, ferromagnetic ordering
of magnetic impurities below $T_{C}$ can greatly enhance the critical current due to an increase
of the pinning force, as seen in conventional superconductors
\cite{Rizz}. This
will sharply increase both the density of the $\pi$ junctions
and the orbital moments just below $T_{C}$. This picture can naturally
explain the result in Fig.~3b. If this interpretaion is correct, our
present results imply ultra-high temperature superconductivity in our MWNT mat
samples, in agreement with theoretical
predictions \cite{Gon,Schr}.

\noindent
{\bf Acknowledgment:} We thank M. Du, G.  Gao and F.  M.  Zhou in 
the 
Department of Chemistry and Biochemistry at CSULA for the elemental 
analyses using ICP-MS. We also thank the Palmdale Institute of Technology for the use of the VSM and 
Lockheed Martin Aeronautics for the cryogens.  This research is partly supported by a 
Cottrell Science Award from Research Corporation.


\begin{thebibliography}{99}
\bibitem{note}The bulk saturation magnetization $M_{s}$ of Co at 1100 K is calculated 
to be 125 emu/g from the value (162 emu/g) of $M_{s}$ at room temperature 
and $T_{C}$ = 1404 K using the
temperature dependence of $M_{s}$ for Ni; For bulk CoFe, $M_{s}$ at 1100 K
is 141 emu/g.
\end{thebibliography}
\end{document}